\documentclass[12pt,preprint]{aastex}
\def\E{\varepsilon}

\shorttitle{$^3$He and $^4$He Acceleration}

\begin{document}


\title{Stochastic Acceleration of $^3$He and $^4$He by Parallel Propagating Plasma Waves}
\vspace{1cm}

\author{Siming Liu\altaffilmark{1}, Vah\'{e} Petrosian\altaffilmark{1, 2}, and Glenn M. 
Mason\altaffilmark{3}}


\altaffiltext{1}{Center for Space Science and Astrophysics, Department of Physics, Stanford
University, Stanford, CA 94305; liusm@stanford.edu}
\altaffiltext{2}{Department of Physics and Applied Physics, Stanford University, Stanford, 
CA 94305; vahe@astronomy.stanford.edu}
\altaffiltext{3}{Department of Physics and Institute for Physical Sciences and Technology, 
University of Maryland, College Park, MD 20742-4111; glenn.mason@umail.umd.edu}


\begin{abstract}

Stochastic acceleration of $^3$He and $^4$He from a thermal background by parallel
propagating turbulent plasma waves with a single power-law spectrum of the wavenumber is
studied. In the model, both ions interact with several resonant modes. When one of these
modes dominates, the acceleration rate is reduced considerably. At low energies, this
happens for $^4$He, but not for $^3$He where contributions from the two stronger modes are 
comparable so that acceleration of $^3$He is very efficient. As a result, the acceleration 
of $^4$He is suppressed by a barrier below $\sim 100$ keV nucleon$^{-1}$ and there is a 
prominent quasi-thermal component in the $^4$He spectra, while almost all the injected 
$^3$He ions are accelerated to high energies. This accounts for the large enrichment of 
$^3$He at high energies observed in impulsive solar energetic particle events. With 
reasonable plasma parameters this also provides a good fit to the spectra of both ions. 
Beyond $\sim 1$ MeV nucleon$^{-1}$, the spectrum of $^3$He is softer than that of $^4$He, 
which is consistent with the observed decrease of the $^3$He to $^4$He ratio with energy.  
This study also indicates that the acceleration, Coulomb losses and diffusive escape of the
particles from the acceleration site {\it all} play important roles in shaping the 
ion spectra. This can explain the varied spectral shapes observed recently by the {\it 
Advanced Composition Explorer}.

\end{abstract}





\section{Introduction}

The abundance and spectra of most ions observed in impulsive solar energetic particle
events (SEPs) qualitatively agree with predictions of stochastic acceleration (SA) by
plasma waves or turbulence (PWT) (e.g. M\"{o}bius et al. 1980; 1982). However, the observed 
enrichment of $^3$He, sometimes by more than four orders of magnitude over its coronal 
abundance (Hsieh \& Simpson 1970; Serlemitsos \& Balasubrahmanyan 1975; Mason et al. 2002), 
remains a theoretical challenge. The aim of this paper is to provide an explanation for 
this observation in the frame of SA models.

Solar $^3$He rich events are enriched also in other heavy elements (Hurford et al. 1975; 
Mason et al. 1986) and are closely related to impulsive X-ray flares, scatter-free electron 
events, and type III radio bursts (Reames, von Rosenvinge \& Lin 1985; Reames, Meyer \& von
Rosenvinge 1994; Mazur, Mason \& Klecker 1995; Mason et al. 2002), indicating that a common
particle acceleration process may be responsible for all these features. Among the
acceleration mechanisms applied to solar flares, the theory of SA, a second order Fermi
acceleration process by PWT, has achieved several significant successes.
The theory was first introduced to account for the acceleration of protons and other ions
responsible for the observed nuclear gamma ray line emissions (Ramaty 1979; Hua, Ramaty \&
Lingenfelter 1989). Later it was shown that it is also applicable to the acceleration of
electrons, which produce the impulsive hard X-ray emission (Miller \& Ramaty 1987; Hamilton
\& Petrosian 1992; Park, Petrosian \& Schwartz 1997; Petrosian \& Donaghy 1999). A recent 
study of SA by cascading Alfv\'{e}n turbulence indicated that it may also explain heavy ion 
acceleration as well (Miller 2003).

It is generally accepted that the enrichment of $^3$He is related to its peculiar
charge to mass ratio so that certain plasma waves resonantly accelerate or heat it to high
energies, producing the observed abundance ratio at a few MeV nucleon$^{-1}$ (M\"{o}bius et
al. 1980; Zhang 1999). Previous work either discusses the feasibility of selective
accelerating $^3$He from very low energies (Temerin \& Roth 1992; Miller \& Vi\~{n}as
1993), or considers a two-stage mechanism; a selective preheating process followed by an
acceleration applicable to all ions (Fisk 1978). The former cannot be compared with
observation directly because the acceleration of other ions has not been addressed
accordingly. For the latter, the observed abundance pattern at high energies is determined
by the preheating processes (Paesold, Kartavykh \& Benz 2003) and the predicted spectrum
depends on several parameters with the corresponding physical processes unspecified 
(Sakurai 1974; Zhang 1995). 

Given its simple dispersion relation, Alfv\'{e}n turbulence is used in most of the SA
models (Ramaty 1979; M\"{o}bius et al. 1982), which cannot explain the varied ion spectral
forms and underestimates the production rate of higher energy particles in disagreement
with the recent observations by the {\it Advanced Composition Explorer} ({\it ACE}; Mason
et al. 2002). However, low energy particles mostly interact with cyclotron and/or whistler
waves. To understand the acceleration of particles from a thermal background, it is
necessary to use the exact dispersion relation (Steinacker et al. 1997). We have carried
out such an investigation recently, which addresses the acceleration of electrons vs
protons by waves propagating parallel to magnetic field lines (Petrosian \& Liu 04, PL04
for short). Besides its achievements in explaining many features of solar flares, we found
that the acceleration of low energy protons can be suppressed significantly relative to the
electron acceleration due to the dominance of the wave-proton interaction by the resonant
Alfv\'{e}n mode.

In this letter, we show that a similar mechanism makes the acceleration of low energy
$^3$He much more favorable relative to that of $^4$He. Because their loss rates are
comparable, such a mechanism basically depletes $^3$He from the thermal background to
$\sim$ MeV energies while most of the $^4$He ions are ``trapped'' at low energies,
resulting a prominent quasi-thermal $^4$He component. The model not only explains the
enrichment of $^3$He at a few MeV nucleon$^{-1}$, but also accounts for the flattening of
the $^3$He spectrum at lower energies (Mason, Dwyer \& Mazur 2000; Mason et al. 2002). It
also predicts that beyond a few MeV nucleon$^{-1}$ the spectrum of $^4$He is always harder
than the $^3$He spectrum. In the next section, we present our SA model. Its application to
SEPs is discussed in \S\ \ref{seps}. In \S\ \ref{sum}, we summarize our results and discuss
their implications.

\section{Stochastic Ion Acceleration}


The theory of SA by parallel propagating waves is explored in detail in PL04. Here, we 
highlight a few key points, which are crucial in understanding the ion acceleration 
process. 

First, a magnetized plasma can be characterized by the ratio of the electron 
plasma frequency $\omega_{\rm pe}$ to gyrofrequency $\Omega_{\rm e}$:
\begin{equation}
\alpha = \omega_{\rm
pe}/\Omega_{\rm e} 
= 3.2 (n_{\rm e}/10^{10}{\rm
cm}^{-3})^{1/2}(B_0/100{\rm \ G})^{-1}\,,
\label{alpha}
\end{equation}
where $n_{\rm e}$ is the electron number density and $B_0$ is the large scale magnetic 
field. To describe the wave modes in the plasma, one also needs to know the abundance of 
ions. In our case, the inclusion of $\alpha$-particle, whose number density is about $10\%$ 
of the proton density, is essential while other heavier elements have little effect on the 
dispersion relation $\omega = \omega(k)$ (Steinacker et al. 1997), where $\omega$ is the 
wave frequency and $k$ is the wavenumber.

\begin{figure}
\plotone{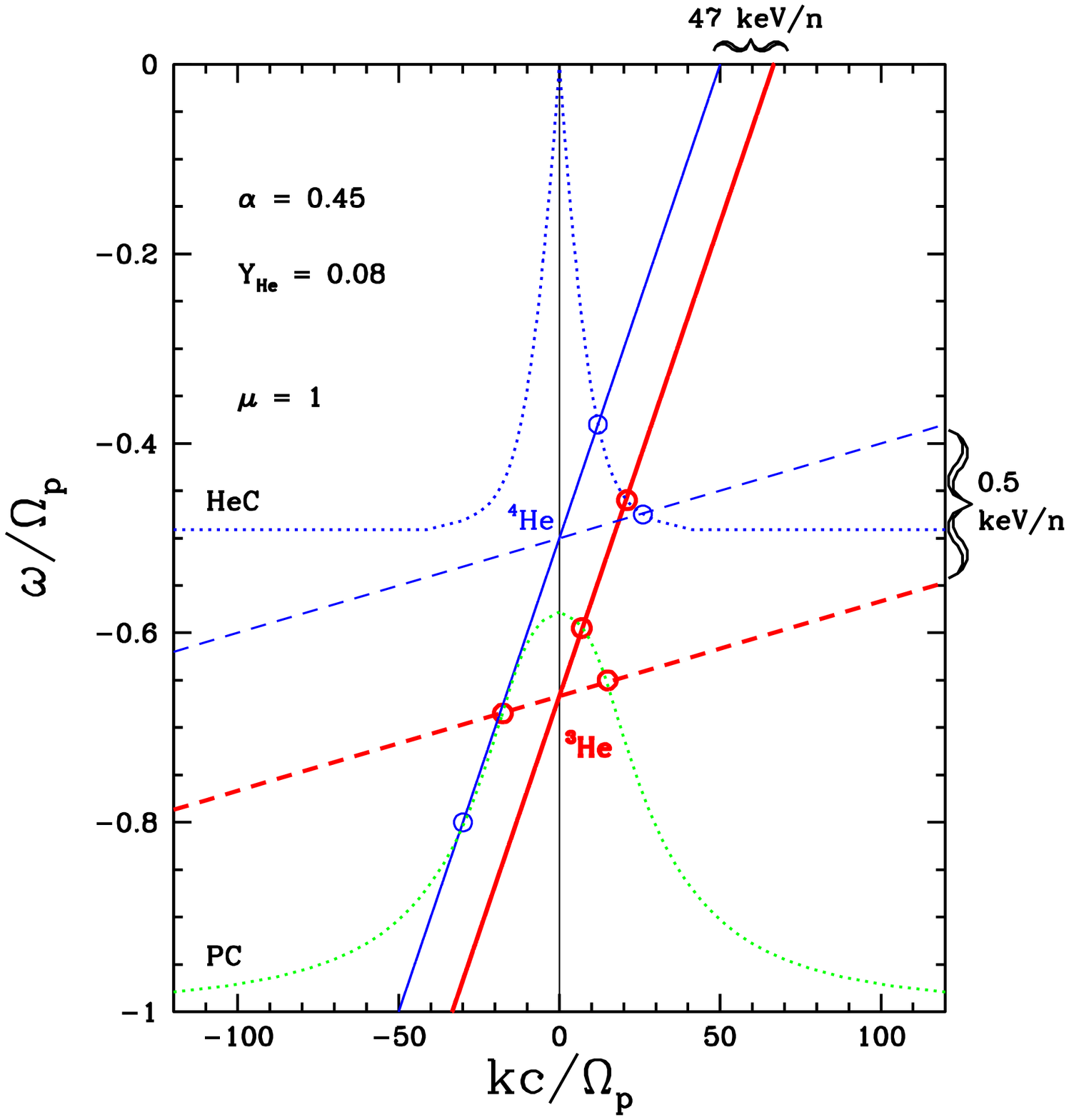}
\caption{
{\it Dotted curves:} dispersion relation of PC and HeC left hand 
polarized waves parallel to the large scale magnetic field.  {\it Solid 
lines:} resonant condition for 47 keV/nucleon $^3$He ({\it thick}) and $^4$He 
({\it thin}). {\it Dashed lines:} resonant condition for 0.5 keV/nucleon. Circles 
designate the resonant points.
\label{fig:disp}
}
\end{figure}

Low energy ions mostly interact with left-handed polarized waves. Figure \ref{fig:disp}
shows the dispersion relation of these wave modes as indicated by the dotted curves in a
plasma with $\alpha=0.45$ and ${\rm Y}_{\rm He} = 0.08$, where ${\rm Y}_{\rm He}$ is the
abundance of $\alpha$-particle (here following the notation in PL04, negative frequencies
indicate that the waves are left-handed polarized). There are two distinct branches, which
we call $^4$He-cyclotron branch (HeC: upper one) and proton-cyclotron branch (PC: lower
one) because they asymptotically approach the corresponding ion cyclotron waves at large 
wavenumbers. At small $k$, HeC branch gives the Alfv\'{e}n waves, while the PC branch 
approaches a wave frequency $\omega_{\rm PC}\simeq(1/2+{\rm Y}_{\rm He})\Omega_{\rm p}$, 
where $\Omega_{\rm p}$ is the proton gyrofrequency.

The acceleration process is determined by the resonant wave-particle interaction. The waves
and particles couple strongly when the resonant condition
$\omega=k\beta c\mu-\Omega_i\gamma^{-1}\,,$ is satisfied, where $\beta c$, $\mu$, 
$\Omega_i$, and $\gamma$ are respectively the velocity, the pitch angle cosine, the 
nonrelativistic gyrofrequency, and the Lorentz factor of the ions.
Figure \ref{fig:disp} also shows this resonant condition (straight lines) for $^3$He and 
$^4$He with $\mu=1$ and two values of energy: $E= 0.5\ {\rm and}\ 47$ keV per nucleon. The 
intersection points of these lines with the dotted curves for the dispersion relation 
satisfy the resonant condition and indicate strong wave-particle couplings. The 
wave-particle interaction rates are calculated by adding contributions from these points 
(PL04).

At low energies (e.g. $E =0.5$ keV nucleon$^{-1}$ in the figure) $^4$He interacts only with
two waves, one from the HeC and one from PC (not shown in the figure because of its large
wavenumber). On the other hand, $^3$He can interact with four waves, three of them 
from the PC branch (two of them are shown in the figure) and one from the HeC branch (also 
not shown in the figure)\footnote{Both ions also interact with a right-handed polarized 
wave in the electron cyclotron branch. Because its contribution to the interaction rates of 
low energy ions is small, we do not discuss this branch here. However, the contribution of 
this wave branch is included in the numerical calculation below.}. As shown in Figure 
\ref{fig:disp} there are two resonant modes for $^3$He with nearly equal and small 
wavenumbers while $^4$He has one resonant mode at low k (see circles in Fig. 
\ref{fig:disp}); the next important resonant mode of $^4$He is at much larger k 
value beyond the range of the figure. Consequently, for a power law spectrum of turbulence, 
the interaction of low energy $^4$He will be dominated by a single mode from the HeC 
branch, giving rise to a very low acceleration rate relative to $^3$He (see discussion 
below)\footnote{It should be noted that this in not true for ions with nearly $90^\circ$ 
pitch angle ($\mu\ll1$). However, this contributes little to the acceleration as a whole 
because there are few such particles}. This is not true for particles with high energies 
($E=47$ keV nucleon$^{-1}$ in the figure). For these particles, the two stronger 
resonant modes have comparable contributions for both ions.

Following previous studies (Dung \& Petrosian 1994; PL04), we will assume that the 
turbulence has a power law spectrum ${\cal E}(k) \propto k^{-q}$ with a low wavenumber 
cutoff $k_{\rm min}$, which corresponds to the turbulence generation scale and determines 
the maximum energy of the accelerated particles. For a turbulence spectral index $q\leq1$, 
a high wavenumber cutoff is also required to ensure the convergence of the total turbulence 
energy density. In this case, we choose $k_{\rm max}$ large enough so that the acceleration 
of particles from the thermal background is not affected by this cutoff. Then the 
characteristic interaction time scale $\tau_{\rm p}$ is given by 
\begin{equation}
\tau_{\rm p}^{-1} = {\pi\over 2}\Omega_{\rm e}\left[{{\cal E}_{\rm tot}\over
B_0^2/8\pi}\right]\cases{(q-1)k_{\rm min}^{q-1},&  for $q>1$; \cr
[\ln{(k_{\rm max}/k_{\rm min})}]^{-1},& for $q=1$; \cr
(1-q)k_{\rm max}^{q-1},& for $q<1$;}
\label{taup}
\end{equation}  
where ${\cal E}_{\rm tot} $ is the total turbulence energy density. 

Ions in a turbulent plasma can gain energy by interacting with the waves with diffusion
coefficients $D_{pp}$, $D_{\mu p}$, and $D_{\mu\mu}$, where $p$ is the momentum of the
particles, and lose energy via Coulomb collisions with electrons and protons in the thermal
background. As a result the particles diffuse in both real and momentum space. The 
spatial diffusion can be approximated with an escape term with a time scale $T_{\rm esc}$. 
Because the Coulomb scattering rate at low energies and the pitch angle scattering rate 
$D_{\mu\mu}$ at high energies are larger than the acceleration rate (except for highly 
magnetized plasmas with $\alpha\ll0.02$), the particle distribution is nearly isotropic 
over all energies. The energy diffusion rate $\gamma(\mu, p) = \tau^{-1}_{\rm ac}(\mu, p) = 
(D_{pp}-D_{\mu p}/D_{\mu\mu})/p^2$ (see Schlickeiser 1989; Dung \& Petrosian 1994). As 
stressed in PL04, when the interaction is dominated by one of the resonant modes, this rate 
becomes negligibly small and the particle acceleration is suppressed. In PL04, we showed 
that such a suppression occurs for low energy protons but not for electrons. 
As mentioned above at low energies the $^4$He interaction is dominated by one mode 
resulting in a low acceleration rate relative to $^3$He. For isotropic pitch angle 
distribution, the particle distribution integrated over the acceleration region, $N(\E)$, 
as a function of the total (not per nucleon) kinetic energy $\E$, satisfies the well known 
diffusion-convection equation:
\begin{equation}
{\partial N\over\partial t}= {\partial^2\over \partial \E^2}(D_{\E\E} N) + 
{\partial\over \partial \E}[({\dot \E}_{\rm L}-A(\E)) N] -{N\over T_{\rm esc}} + Q\,, 
\label{dceq}   
\end{equation}
where $D_{\E\E}=\E^2<\gamma(\mu, p)>\sim \E A(\E)$ is the pitch angle averaged diffusion 
rate, $Q$ is a source term, and the loss rate ${\dot \E}_{\rm L}$ is given in PL04.

Figure \ref{fig:times} shows the relevant time scales for a model with $L = 10^{10}$ cm,
$B_0=500$ Gauss, $n_{\rm e}=5\times 10^9$ cm$^{-3}$ and temperature $T=0.2$ keV. The
turbulence spectral index $q=1$, $\tau_{\rm p}^{-1} = 3$ s$^{-1}$ and $\alpha = 0.45$. 
The acceleration and loss times are defined as: $\tau_{\rm a} = \E/A(\E)\,, \tau_{\rm loss} 
= \E/\dot{\E}_{\rm L}\,.$ Depending on $L$, $\tau_{\rm p}$ and $\E$, either the transit 
time $\tau_{\rm tr}=(L/ \sqrt{2}\beta c)$ or the diffusion time $\tau_{\rm dis}= 2\tau_{\rm 
tr}^2/\tau_{\rm sc}$ can dominate the escaping process, where $\tau_{\rm sc}$ is the 
scattering time including both wave-particle scatterings and particle-particle collisions. 
We define $T_{\rm esc} = (L/ \sqrt{2}\beta c)(1+{\sqrt{2}L/\beta c \tau_{\rm sc}})\,,$ 
which incorporates both processes of escape. 

There are several features in these times, which require particular attention:

First, all the time scales are energy dependent and the acceleration times have 
complicated forms, in contrast to the often used simplified acceleration by pure Alfv\'{e}n 
turbulence and/or with energy independent escape time. This affects the spectral shapes 
significantly. 

Second, the typical time scale varies from less than one second to hundreds of seconds with 
different processes dominating at different energies.

Third, the acceleration times for $^4$He and $^3$He are quite different due to interaction 
with different waves, while their escape and loss times are similar because these times are 
dominated by Coulomb collisions and wave-particle scatterings ($\propto <D_{\mu\mu}>$). 

Fourth, the acceleration time of $^3$He is shorter than the Coulomb loss time below a few 
hundreds of keV nucleon$^{-1}$. Consequently, for a background plasma of a few million 
Kelvin almost all the injected $^3$He can be accelerated to high energies (a few MeV 
nucleon$^{-1}$). Above this energy the escape time becomes shorter than the acceleration 
time and the acceleration is quenched.

Fifth, $^4$He acceleration time increases sharply with the decrease of energy near 100 
keV nucleon$^{-1}$. This is because at lower energies the wave-particle interaction is 
dominated by one of the resonant modes. $^4$He acceleration time starts to decrease 
with energy near 50 keV nucleon$^{-1}$ where the two resonant waves have comparable 
wavenumber (Figure \ref{fig:disp}). Because the acceleration time is longer than the 
loss time at low energies, for an injection plasma with a temperature less than 
$\sim 10^8$ K, only a small fraction of $^4$He in the Maxwellian tail can be accelerated to 
very high energies. The rest is heated up into a quasi-thermal distribution.

Sixth, the acceleration time of $^4$He is shorter than that of $^3$He above $\sim 100$ keV 
nucleon$^{-1}$, so that the $^4$He acceleration becomes more efficient at these energies.

\section{Application to SEPs}
\label{seps}

Because the duration of SEPs are much longer than the characteristic time scales 
discussed in the previous section, we will assume that the system reaches a steady state. 
One can then ignore the time derivative term in equation (\ref{dceq}) and solve this 
equation for the escaping flux $f = N/T_{\rm esc}$, which can be directly compared with the 
observed fluences.

\begin{figure}
\plotone{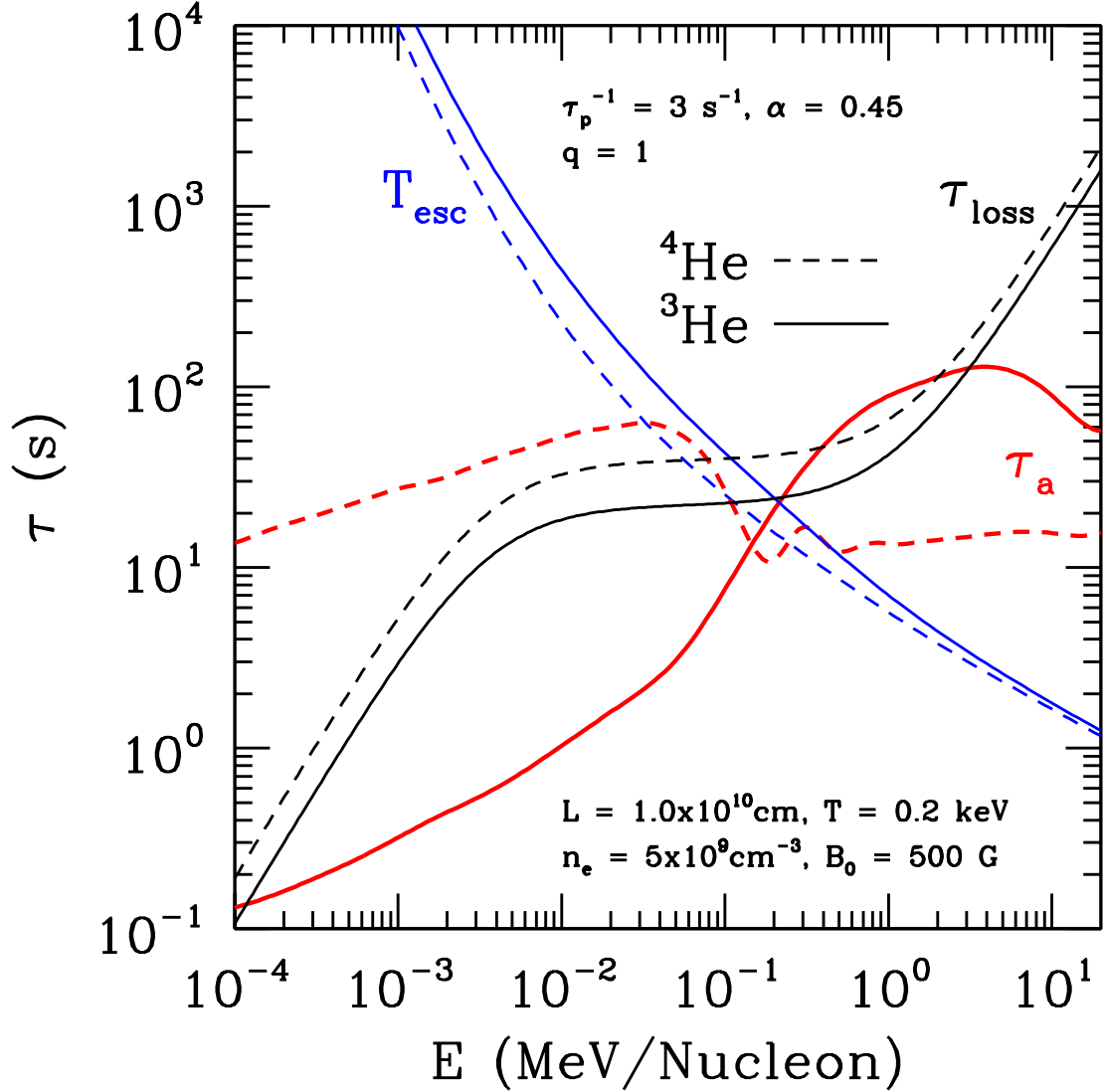}
\caption{
Time scales of a SA acceleration model. The model parameters are indicated in the figure. 
{\it Solid:} $^3$He, {\it dashed:} $^4$He. The thicker curves indicate the 
acceleration times. See text for details.
\label{fig:times}}
\end{figure}

\begin{figure}
\plotone{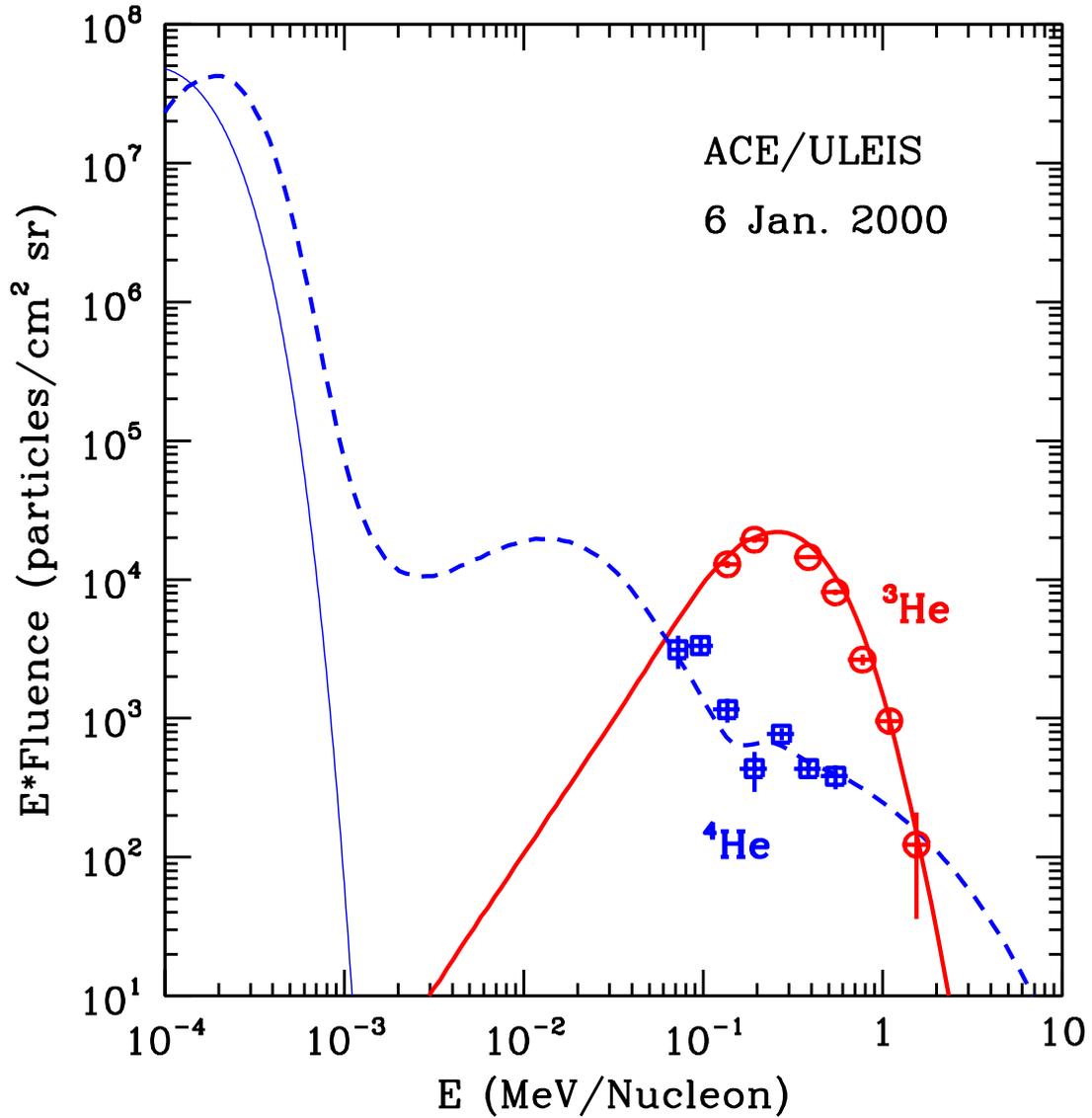}
\caption{
Model fit to the spectra of the SEP on Jan. 6, 2000. The thick solid curve is for $^3$He 
and the dashed curve is for $^4$He. The injected particles have a temperature of $T = 0.2$ 
keV and the corresponding distribution of $^4$He is indicated by the thin solid curve with 
arbitrary normalization. All other model parameters are shown in Figure \ref{fig:times}.
\label{fig:spec1}
}
\end{figure}

Figure \ref{fig:spec1} shows our model fit to the spectra of an SEP event on January 6, 
2000, the event with the largest $^3$He enrichment (Mason et al. 2000). The data are 
obtained by carefully subtracting the background and avoiding contaminations from nearby 
events. For the model spectra, we assume that the injected plasma has a coronal abundance, 
i.e. the abundance of $^4$He is two thousand times higher than that of $^3$He. The model 
parameters are the same as those in Figure \ref{fig:times}. The temperature of 0.2 keV is a 
typical value for impulsive SEP events (Reames et al. 1994). Because the acceleration 
region is connected to open magnetic field lines probably above the flaring coronal loops, 
a size of $L=10^{10}$cm seems quite reasonable. The magnetic field and gas density are also 
characteristic of an upper coronal reconnection region. The turbulence parameters $q$ and 
$\tau_{\rm p}$ are free model parameters.

The model not only accounts for the enrichment of $^3$He, but also gives a reasonable fit
to the spectra. Because at low energies the acceleration time of $^3$He is shorter than any
other times, almost all of the injected $^3$He ions are accelerated to a few hundreds of
keV nucleon$^{-1}$, where $\tau_{\rm a}$ becomes longer than the loss and escape time,
resulting in a sharp cutoff. The $^4$He acceleration time, because of the barrier mentioned
above, is much longer than its loss time below a few keV nucleon$^{-1}$. This suppresses
the $^4$He acceleration so that most of the injected $^4$He ions remain near the injection
energy and form a quasi-thermal component. However, even when $A(\E)\le\dot{\E}_{\rm L}$
and the direct acceleration is difficult, some $^4$He ions can still diffuse to high
energies due to the first term on the right hand side of equation (\ref{dceq}).
Consequently, we expect a nonthermal tail in addition to the quasi-thermal component. This
tail is cut off at $\sim 10$ MeV nucleon$^{-1}$ (Fig. \ref{fig:spec1}), when the escape
time becomes much shorter than the acceleration time (Fig. \ref{fig:times}). The
acceleration barrier extends to $\sim 100$ keV nucleon$^{-1}$, most of the $^4$He ions
diffuse to higher energies are stopped below this energy. About $10\%$ of the particles can
be accelerated to even higher energies and produce the observed high energy spectrum with
some spectral features in this energy range. However, a more realistic model, including
acceleration by other wave modes, is expected to produce smoother spectra.

\begin{figure}
\plotone{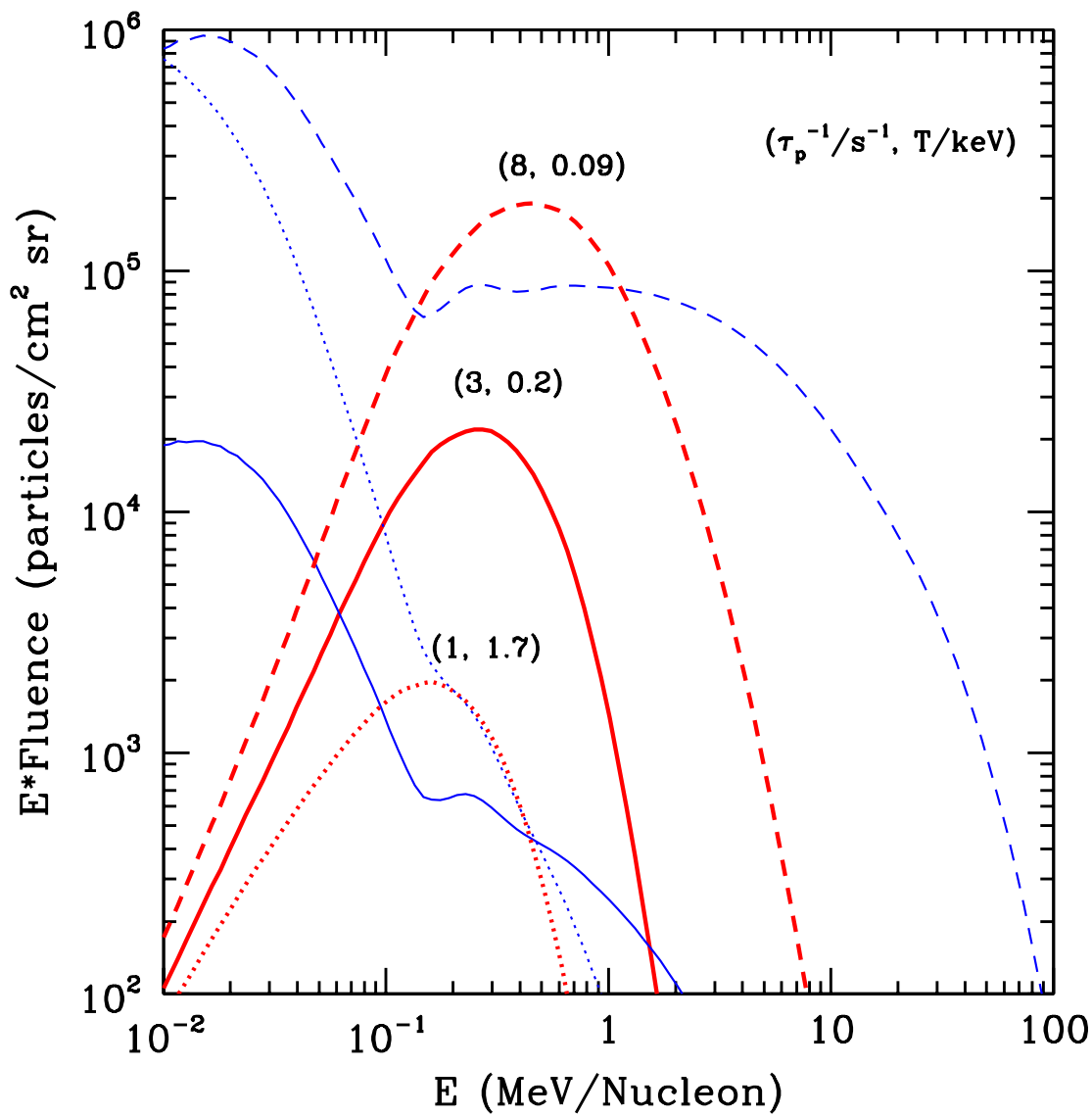}
\caption{
\label{fig:spec2}
Dependence of the $^3$He (thick curves) and $^4$He (thin curves) spectrum on $\tau_{\rm 
p}$ and $T$. The dotted, solid, and dashed curves are for $\tau_{\rm p}^{-1}$ = 1, 3, and 8 
s$^{-1}$ and $T$ = 1.7, 0.2, and 0.09 keV, respectively. The dotted (dashed) curves are 
shifted up (down) by decade. All other model parameters remains the same as those in Figure 
\ref{fig:times}. 
}
\end{figure}

The above behaviors depend on six model parameters, which are shown in Figure
\ref{fig:times} (note that $\alpha$ is given by $n_{\rm e}$ and $B_0$ via
eq.[\ref{alpha}]). Figure \ref{fig:spec2} shows how the spectra change with two of these
parameters: $\tau_{\rm p}^{-1}$ (or the energy density of the turbulence) and the injection
temperature $T$. Because of the acceleration barrier at low energies, the acceleration of
$^4$He is very sensitive to $T$. In the figure, we adjust $\tau_{\rm p}$ in order to make
the fluence of $^4$He and $^3$He comparable at a few hundred keV per nucleon. As evident,
$^3$He spectrum peaks at a higher energy with the increase of the intensity of the
turbulence and the $^3$He enrichment also changes significantly. This can explain the 
variation of the $^3$He spectral peak and its enrichment in class 2 SEPs (Mason et al. 
2000; Mason, Mazur \& Dwyer 2002).

\section{Summary and Discussion}
\label{sum}

In the paper, we have shown that the observed $^3$He and $^4$He spectra in some impulsive
SEPs can be produced via SA by PWT propagating parallel to magnetic field lines. In the
model, the particles are injected from a thermal background. Because the interaction of low
energy $^4$He is dominated by one of the resonant wave modes, an acceleration barrier
forms, giving rise to a prominent quasi-thermal component in the steady state particle
spectrum. This is not true for $^3$He ions, however, which can be efficiently accelerated
by two waves in the PC branch, from very low energies to $\sim$ MeV nucleon$^{-1}$, where
they escape quickly from the acceleration site, producing a sharp spectral cutoff.  This
explains the $^3$He enrichment and its varied spectrum seen in class 2 SEPs, 
which have relatively short rise-time and are more likely associated with the impulsive 
phase particle acceleration directly. Class 1 events have less $^3$He enrichment and 
similar spectra for most ions, indicating a different acceleration process (Mason et al. 
2000). The feasibility of producing such events by shocks or turbulence with a different 
spectrum than that of the present model will be investigated in the future.

Because the $^4$He acceleration time is shorter than that of $^3$He at higher energies and
their loss and escape times are comparable, the model predicts a decrease of the $^3$He to 
$^4$He ratio with energy beyond $\sim 1$ MeV nucleon$^{-1}$, which is consistent 
with observations (Mason et al. 2002; Reames et al. 1997; M\"{o}bius et al.  1982;  
M\"{o}bius et al. 1980; Hsieh \& Simpson 1970). Spectral information at higher energies
is critical in testing this model prediction. {\it We emphasize here that a special 
injection mechanism is not required to explain the observations.} However, a second-phase 
shock acceleration (Van Hollebeke, McDonald \& Meyer 1990) can modify the ion spectra and 
may explain events where the high energy spectral indices of $^3$He and $^4$He are similar
(Serlemitsos \& Balasubrahmanyan 1975; Reames et al. 1997).

The model can also be applied to acceleration of heavy-ions. In this case, the acceleration 
is dominated by resonant interaction with waves in the HeC branch and ions with low charge 
to mass ratio are accelerated more efficiently, which may explain the increase of 
ion enrichment with the increase of their charge to mass ratio (Mason et al. 1986). Because 
$^3$He acceleration is dominated by interaction with the PC branch, which may be decoupled 
from the HeC branch, the model may also explain the lack of quantitative correlation of the
enrichment of $^3$He and heavy-ions (Mason et al. 1986). We will investigate these 
aspects in future work.

The primary uncertainties of the model are associated with the turbulence generation 
mechanism and the subsequent cascade and dissipation processes. We start the investigation 
with a power law spectrum of the wavenumber for the turbulence, which can be characterized 
by the turbulence energy density and the spectral index. The acceleration of particles by 
obliquely propagating waves and turbulence with complex spectral shapes may give 
quantitatively different results. However, the idea of an efficient depletion of low energy 
$^3$He from the source region to high energies to produce the observed enhancement seems well 
founded.

\acknowledgments

This research was partially supported by NSF grant ATM-0312344, NASA grants NAG5-12111, 
NAG5 11918-1 at Stanford and NASA grant PC 251429 at University of Maryland.

\clearpage



\clearpage



\clearpage




\end{document}